\documentstyle[aps,twocolumn]{revtex}

\newcounter{defcount}
\newenvironment{definition}%
  {\stepcounter{defcount}\par\noindent{\bf Definition \arabic{defcount}:}}%
  {\par\bigskip}

\def\opone{\leavevmode\hbox{\small1\kern-3.8pt\normalsize1}}
\def\<{\langle}
\def\>{\rangle}

\def\opone{\leavevmode\hbox{\small1\kern-3.8pt\normalsize1}}
\newcommand{\beq}{\begin{equation}}
\newcommand{\eeq}{\end{equation}}
\newcommand{\beqa}{\begin{eqnarray}}
\newcommand{\eeqa}{\end{eqnarray}}

\newcommand{\Sen}{S}
\def\A{\cal A}
\newcommand{\RO}{{R_0}}
\newcommand{\RI}{{R_1}}
\newcommand{\RP}{{R_p}}
\def\incircle#1
   {\unitlength=1.00mm
   \begin{picture}(3.00,3.00)
   \put(1.00,1.00){\circle{3}}
   \put(1.00,1.00){\makebox(0,0)[cc]{#1}}
   \end{picture}}

\begin{document}

\title{A Quantum solution to the Byzantine agreement problem}
\author{Matthias Fitzi$^1$, Nicolas Gisin$^2$, Ueli Maurer$^1$}
\address{$^1$Department of Computer Science, Swiss Federal Institute of
                     Technology (ETH), CH-8092 Zurich, Switzerland\\
         $^2$Group of Applied Physics, University of Geneva,
                     CH-1211 Geneva 4, Switzerland}

\date{\today}
\maketitle

\begin{abstract}
We present a solution to an old and timely problem in distributed computing. Like
Quantum Key Distribution (QKD), quantum channels make it possible to achieve taks
classically impossible. However, unlike QKD, here the goal is not secrecy but
agreement, and the adversary is not outside but inside the game, and the resources
require qutrits.
\end{abstract}

\vspace{7mm}

Entanglement is a resource that allows quantum physics to perform tasks
that are classically impossible. This is the new leitmotif of quantum
information processing. The best known examples are quantum
cryptography\cite{BB84,Ekert91} and Shor's algorithm to
efficiently factorize large numbers\cite{Shor94}. There are numerous
other examples, but admittedly many of them look quite artificial. This
can be understood by the fact that the field is rather young and one has
to learn to think about problems in a completely new way. In this letter
we consider an old information-theoretical problem in the field of
fault-tolerant distributed computing, known as 
{\it the Byzantine agreement problem}~\cite{LaShPe82} and present a solution which
exploits entanglement between three qutrits (i.e., three $3$-dimensional
quantum systems). 

Imagine that several divisions of the Byzantine army are camped outside an
enemy city, each division commanded by its own general. The generals can
communicate with one another only by messengers. One of the generals, the
commanding general, after observing the enemy, must decide on a plan of
action and communicate it to the other generals.
However, some of the generals (especially the commanding general himself)
might be traitors, trying to prevent the loyal generals from reaching
agreement on the plan of action. The question hence is whether there is a
protocol among the generals that, after its termination, satisfies the
following conditions:
\begin{enumerate}
 \item All loyal generals agree on a common plan of action.
 \item If the commanding general is loyal, then all loyal generals agree
  on the commanding general's plan.
\end{enumerate}

More precisely we define Byzantine agreement (shortly broadcast) as follows.

\medskip

\begin{definition}\label{def:bc}
 A protocol among $n$ players such that one distinct player $\Sen$
 (the sender) holds an input value $x_S\in{\cal D}$ (for some finite
 domain ${\cal D}$) and all other players (the receivers) eventually
 decide on an output value in ${\cal D}$ is said to achieve {\em broadcast}
 if the protocol guarantees that all honest players decide on the same
 output value $y\in{\cal D}$ and that $y=x_S$ whenever the sender is honest.
\end{definition}

In modern terms, this problem concerns coordination in distributed computing
(among several processors or computers) where some of the processors might fail.
For example, a database can be replicated among several servers in order to
guarantee access to the database even if some of the servers misbehave.
Nevertheless, an inconsistent external update of the database must result in
all honest servers having exactly the same views on the database.
Consider for instance that the database contains the price of valuable goods,
or currency exchange rates. It is then important that no adversary, not even an inside adversary, can
affect the coordination in such a way that the prices would be low somewhere and
high elsewhere.

The broadcast problem has been considered in a vast literature and has
developed several variations \cite{LaShPe82}.
Here we shall define the problem more precisely as follows. Three players
are connected by pairwise authenticated classical and quantum channels, see Fig. 1.
For simplicity, we assume the channels to be error free --- generally, errors
would have to be additionally dealt with by means of error correction codes.
The general purpose is that one of the players, namely the sender 
($\Sen$ for short), broadcasts a bit to his two partners, the receiving players
$\RO$ and $\RI$.
Both receivers should end with the same value. However, one -- and at most
one -- of the three players might actually be an active adversary. For instance,
a dishonest sender could send different bit values to $\RO$ and $\RI$. The
receivers may realize that there is a problem simply by exchanging their bits.
But then, player $\RI$ cannot conclude whether the sender is dishonest and he
should keep the bit received from $\RO$ or whether player $\RO$ is cheating
and he should keep the bit received from the sender.
It is not too difficult to convince oneself that because the players have
only access to pairwise channels, this task is not obvious. 

If the players have access only to classical pairwise authenticated channels, the
broadcast problem is provably unsolvable~\cite{LaShPe82,FiLyMe86}. This even
holds for arbitrary pairwise communication, i.e., not even quantum channels
can help to solve the problem~\cite{FGMO01}.
However, we shall demonstrate that the additional resource of the quantum
channel allows them to solve a slightly weaker problem, namely detectable
broadcast, which is powerful enough for a large range of applications of this
problem.

\medskip

\begin{definition}\label{def:dbc}
 A protocol among three players such that one distinct player $\Sen$
 (the sender) holds an input value $x_S\in{\cal D}$ (for some finite
 domain ${\cal D}$) and the other two players (the recipients) eventually
 decide on an output value in ${\cal D}$ is said to achieve
 {\em detectable broadcast} if the protocol satisfies the following
 conditions:
 \begin{enumerate}
  \item If no player is corrupted then the protocol achieves broadcast.
  \item If one or more players are corrupted then either the protocol
   achieves broadcast or all honest players abort the protocol.
 \end{enumerate}
\end{definition}

Note that detectable broadcast cannot be solved only with pairwise
authenticated classical channels. However, we demonstrate that pairwise
authenticated classical {\it and quantum} channels are sufficient to
solve the problem. Basically, we solve the problem by having the players
\begin{enumerate}
 \item distribute entanglement,
 \item check that the entangled states are not corrupted, and
 \item use them to solve the problem. 
\end{enumerate}

At first sight, this sounds very similar to quantum cryptography.
But, actually, it is very different! Indeed, here we do not require any
secrecy: what counts is to avoid any discord. 
Also, here, in contrast to quantum cryptography, the adversary
is not an outside player, but might be anyone among the three players.
Finally, here, the entanglement requires entangled qutrits, contrary to quantum
cryptography where qubits suffice.

The first point of the above program, i.e. distributing entanglement, is
trivial (in theory), since we assume that quantum channels are available.
The testing (i.e.~the second point), however, is tricky. Indeed, the
testing requires (classical) communication between the three players. But
the adversary being inside the game could corrupt this communication phase.
Especially at the last round of the communication phase, the adversary
could send contradictory messages to the two honest players. In other
quantum information protocols involving more than two players, e.g., in
quantum secret sharing~\cite{qss}, this problem is avoided by assuming that
the players can broadcast their (classical) messages. But here broadcasting
is not assumed among the primitives, on the contrary, it is the goal of the
game. Below we show how to break this vicious circle! But first, we need to explain 
how the three players can use entangled qutrits to solve the problem.

Let us assume that the $3$ players share many qutrits triplets $\Psi_j$,
each in the Aharonov state $|\A\>$:
\beqa
|\A\>&=&\big(|0,1,2\>_{\vec m}+|1,2,0\>_{\vec m}+|2,0,1\>_{\vec m}\label{Astate}\\
&-&|0,2,1\>_{\vec m}-|1,0,2\>_{\vec m}-|2,1,0\>_{\vec
m}\big)\frac{1}{\sqrt{6}} \nonumber
\eeqa
where $|0,1,2\>_{\vec m}$ denotes the tensor product state 
$|0\>_{\vec m}\otimes|1\>_{\vec m}\otimes|2\>_{\vec m}$.
If one identifies qutrits with spin-1 and associates the state $|2\>_{\vec m}$ with
the eigenvalue $-1$ of the spin operator $\vec m \vec S$, then the Aharonov state is
the unique three spin-1 state of total spin zero. Consequently -- and
analogously to the singlet state of qubit pairs -- the Aharonov state is
invariant under tri-lateral rotations: it keeps the same form (\ref{Astate})
for all directions $\vec m$, where $|0\>_{\vec m}$, $|1\>_{\vec m}$ and
$|2\>_{\vec m}$ are the 3 eigenvectors of the spin operator $\vec m\vec S$.
We shall exploit the fact that whenever the three qutrits are all measured
in the same basis, then all three results differ.

With the help of this additional resource, i.e., the Aharonov states, the
protocol runs as described below. At each step we comment on the reasons why
this is safe. Actually, all steps are rather trivial, except the last one
which needs a careful analysis.
\begin{enumerate}
\item First, the sender $\Sen$ sends the bit $x$ to be broadcast to the two
receivers $\RO$ and $\RI$, using the classical channels. Let us denote $x_0$
and $x_1$ the bits received by $\RO$ and $\RI$, respectively. Next, the Sender
$\Sen$ measures all his qutrits in the z-basis.
Whenever he gets the result $x$, $\Sen$ sends the index $j$ to both
receivers\cite{inpractice}.
Accordingly, the players $\RO$ and $\RI$ receive each a set of indices,
$J_0$ and $J_1$, respectively (label \incircle{1} ~on Fig. 1).

\item Both receivers test the consistency of their data. For this they
measure their qutrits in the z-basis. If all results with indices in $J_p$
differ from $x_p$, then player $\RP$ has consistent data and he sets a
flag $y_p=x_p$. If a set of data is inconsistent, then the player sets his
flag to $y_p=\perp$ (interpreted as {\it inconsistent}).

\item The two receivers send their flags to each other. If both flags agree
then the protocol terminates with all honest players agreeing on $x$.

\item If $y_p=\perp$, player $\RP$ knows that the sender is dishonest. He
concludes that the other receiver is honest and he simply accepts the bit
he receives from him (If $y_0=y_1=\perp$, then they both end with the
"value" $\perp$).

\item It remains only the interesting case that both receivers claim that
they received consistent, but different, data. The strategy we propose
then is that player $\RI$ will not change his bit $y_1$, unless player $\RO$
convinces him that he did indeed receive the bit $y_0$ from the sender in a
consistent way. To convince his partner of his honesty, player $\RO$ sends
him all the indices $k\in J_0$ for which he has the result $1-y_0$ (label \incircle{2} ~on Fig. 1).

\item Receiver $\RI$ now checks that he gets "enough" indices $k$ from
 $\RO$ such that
 \begin{enumerate}
  \item "almost all\cite{almostall}" indices $k$ from $\RO$ are not in $\RI$'s index set
   $J_1$, and such that
  \item these $k$ indices correspond to qutrits for which $\RI$'s results
   are "almost all" equal $2$.
 \end{enumerate}
If $\RO$ indeed got an index set that is consistent with bit $y_0$ then
$\Sen$ holds $y_0$, $\RO$ holds $1-y_0$, and hence, $\RO$'s result must
be a $2$. If the test succeeds, player $\RI$ changes his bit to $y_0$,
otherwise he keeps $y_1$.
\end{enumerate}

Let us examine why player $\RO$ cannot cheat (see Table 1).
Assume that $\RO$ receives the bit $x_0=0$, but pretends that he got $1$.
To convince the receiver $\RI$ to accept his "pretended bit", player $\RO$
must first announce that he received consistent data (which is true, but
with bit value $0$), and next send a sufficiently large set of indices
$\{k\}$ with almost no intersection with $J_1$ and for which $\RI$ almost
always has the result $2$. Since player $\RO$ has no information on the
indices outside $J_0$ for which he measured $1-y_0$ (other than $1-y_0$
itself), approximately half of the indices that player $\RI$ gets are
different from $2$ --- which is not accepted by player $\RI$.

Let us stress two important features of this protocol. First, the last
player, i.e., player $\RI$, almost never talks (he only sends his value
$y_1$ to $\RO$). Moreover, if the sender is honest, then the last player
never changes his mind. This is important for the distribution and test
phase of the protocol described below.
Second, a third noticeable point is that our protocol uses trits and not
bits. Clearly, trits can always be encode as 2 bits, like qutrits can be
realized as $2$-qubit in symmetric states. But fundamentally, the protocol
requires trits (and we convinced ourselves, but without a proof, that no
protocol using intrinsically only bits exists).

So far we described a protocol assuming that the 3 players
share a large collection of qutrit triplets in the Aharonov state
(\ref{Astate}). We now describe a protocol to distribute and
test such states. This protocol uses only pairwise communication, in
particular no broadcasting is assumed. Nevertheless, the protocol has only
two possible outcomes: global success or global failure. By global we mean
that all honest players end with the same conclusion. In case of failure,
the broadcasting protocol does not even start. In case
of success, broadcasting can be realized reliably.

The distribution-\&-testing protocol works as follows:
\begin{enumerate}
\item Player $\RI$ prepares many qutrit triplets $\Psi_j$ in the Aharonov
 state (\ref{Astate}). For each index $j$ he sends one qutrit to player
 $\Sen$ and one to $\RO$ (label \incircle{A} ~on Fig. 1).

\item Both $\Sen$ and $\RO$ check that their qutrits are in the maximally
 mixed state. In case of success, they set a flag $f_p$ to 1, else to 0.

\item Player $\RO$ sends a sample of his qutrits to
 $\Sen$ (label \incircle{B} ~on Fig. 1). 
 Player $\Sen$ tests that the sample of qutrit pairs he now holds
 are in the correct state\cite{Atest}:
 \beqa
  \rho_{s_{\RO}}&=&Tr_{,\RI}(|\A\>\<\A|)\\
   &=&\frac{1}{3}\left(P_{|1,2\>-|2,1\>}+P_{|2,0\>-|0,2\>}+P_{|0,1\>-|1,0\>}\right)
   \nonumber\label{rho12}
 \eeqa

 If the test fails,  then he sets his flag to 0: $f_{\Sen}=0$.

\item Player $\RI$ sends a sample of his qutrits to $\RO$ and another sample
 to $\Sen$. Both $\RO$ and $\Sen$ test their qutrit pairs as in the previous point 3.
 If the test fails they set their flag to 0.

\item Player $\Sen$ and $\RO$ exchange their flag. If a player receives a 0, then he
 sets his flag to zero.

\item Both players $\Sen$ and $\RO$ broadcast their flags using the
 protocol described previously.

\item Any player with flag $1$ who received a $0$-flag changes his flag
 to $0$.
\end{enumerate}
At first, the step 6 of the above protocol seems impossible, since the
broadcast protocol requires reliable Aharonov states! Nevertheless, let us
look closer at this step. If player $\RI$ does not produce the correct
states, then, since by assumption there is no more than one dishonest player,
players $\Sen$ and $\RO$ are honest and both will end with their flag on
failure: $f_{\Sen}=f_{\RO}=0$. Let us thus assume that all states
$\Psi_j=|\A\>$. Consequently, the broadcasting is reliable. All what a
dishonest player $\Sen$, $\RO$ or $\RI$ could do is to act in such a way that the
flags are set to $0$ \cite{blocking}. But during the last step of the protocol, i.e.,
the broadcast sessions, both the one initiated by $\Sen$ and the one
initiated by $\RO$ are reliable. Hence it is impossible that some players
end this protocol thinking that a status of success has been reached, while
another one thinks the opposite. Moreover, if all players agree on success,
then they share Aharonov states and they can reliably run the broadcast
protocol.

The field of quantum communication is still in its infancy. Only very few
protocols concern more than two parties and almost all use qubits. In this
letter we presented a protocol among three players connected by pairwise
quantum channels able to transmit qutrits and to preserve their entanglement.
The protocol is a version of the well known Byzantine agreement problem, a
very timely problem in today's information based society. Admittedly, the
problem has been slightly adapted to fit into the quantum frame, a natural
synergy between classical and quantum information theories. It is not too
difficult to generalize our result to $n$ players with $t<\frac{n}{2}$
cheaters, though this is beyond the present letter\cite{FGMR01}. One may question whether
the use of qutrits is necessary or not for broadcasting. Though this is still
an open question, it is clear that the present protocol is intimately related
to the Aharonov state (the natural generalization to three parties of the
well known singlet state of two qubits), hence to qutrits.

Two other features of our protocol should be mentioned. First, the quantum states
are used "only" to distribute classical private random variables with specific correlation
to the 3 players (a trit per player, each
of a different value, all combination with equal probabilities). This is similar to
quantum cryptography where quantum mechanics "only" provides key distribution. However,
contrary to the "one-time-pad" algorithm used in conjunction with quantum key distribution,
the present algorithm was itself inspired by the elegance of the Aharonov quantum state.
Finally, experimental demonstration of the protocol can be realized with today's technology,
using photons and 3-paths interferometers. Actually one would not need to prepare 3
entangled photons, two would suffice since the preparer $\RI$ could measure his qutrit
immediately, similarly to the demonstration of quantum secrete sharing using pairs
of photons \cite{QSSTittel}.\\

Work {\bf supported} by the Swiss Science Foundation.



\section*{Figure Captions}
\begin{enumerate}
\item{Fig 1: Flow of classical (straight lines) and quantum (wavy lines) information.
Note that both kinds of information flow exactly in opposite direction. This is needed to
avoid that the adversary can bring in confusion at the last communication round.}
\end{enumerate}

\begin{table}
\centering
$\matrix{ & \matrix{I & II}  & \matrix{III & IV}  & \matrix{V & VI} \cr S &
\matrix{0 & 0}  & \matrix{1 & 1}  & \matrix{2 & 2} \cr R_{0} & \matrix{1 & 2}  &
\matrix{2 & 0}  & \matrix{0 & 1} \cr R_{1} & \underbrace {\matrix{2 & 1}
}_{J_{0}} & \underbrace {\matrix{0 & 2} }_{J_{1}} & \matrix{1 & 0} }$ 
\caption{After measuring their qutrits, the sender $\Sen$ and receivers $\RI$ and
$\RO$ results fall into 6 classes, labeled with Roman numbers. The index-sets $J_0$
and $J_1$ associated to the bit values 0 and 1 correspond to the labels I,II and 
III,IV, respectively. If $\RO$ receives the bit 0 and the set $J_0$ he can announce to $\RI$ all
cases where he has the bit 1: all cases labeled by I. For all these cases $\RI$ has the
value 2. However, if $\RO$ tries to cheat and pretends to have received a bit 1, then he can't
differentiate between the cases labeled IV and V. For the latter, $\RI$ has a value 1, he can
thus detect cheating $\RO$.}
\end{table}


\begin{thebibliography}{90}

\bibitem{BB84} C. Bennett, G. Brassard, Int. conf. Computers, Systems \& Signal
Processing, Bangalore, India, December 10-12, 175-179, 1984.

\bibitem{Ekert91} A. Ekert, Phys. Rev. Lett. {\bf67}, 661, 1991.

\bibitem{Shor94} P. W. Shor, {\it{Proceedings of the 35th Symposium
on Foundations of Computer Science}}, Los Alamitos, edited by
Shafi Goldwasser (IEEE Computer Society Press), 124-134, 1994.

\bibitem{LaShPe82}
L. Lamport, R. Shostak, and M. Pease.
\newblock The {B}yzantine generals problem.
\newblock {\em {ACM} Transactions on Programming Languages and Systems},
  4(3):382--401, July 1982; and refs therein.

\bibitem{FiLyMe86}
M.~J. Fischer, N.~A. Lynch, and M.~Merritt.
\newblock {\em Distributed Computing}, 1:26--39, 1986.

\bibitem{FGMO01}
M.~Fitzi, J.~A. Garay, U.~Maurer, and R.~Ostrovsky.
\newblock {\em Advances in Cryptology - CRYPTO '01, Lecture Notes in Computer Science},
          Springer-Verlag. 2001.

\bibitem{qss} M. Hillery, V. Buzek, and A. Berthiaume, Phys. Rev. A {\bf59}, 1829, 1999;
A. Karlsson, M. Koashi, and N. Imoto. Phys. Rev. A {\bf59}, 162, 1999;
R. Cleve, D. Gottesman, H.-K. Lo, Phys. Rev. Lett. {\bf 83}, 648, 1999.

\bibitem{inpractice} In practice, it is more realistic to assume that each
player measures all the qutrits in randomly chosen bases. Then, each time
they communicate a trit value, they need to add the information about the
measurement basis, and each time they receive a trit they ignore it unless
it happens that they measured their corresponding qutrit in the same
basis.

\bibitem{almostall} This qualification is needed for statistical tests. In the limit of
arbitrarily many qutrit-triplets, "almost all" translates into "with probability one".

\bibitem{Atest} To see that this test is sufficient to
 guarantee that the player $\RI$ who prepared the states can't cheat,
 consider a general purification $\Psi$ of the mixed state $\rho_{s_{\RO}}$.
 Dividing the three parts into $\RI$ versus the two others, one can write
 $\Psi$ in the Schmidt form. Using the fact that the eigenstates of
 $\rho_{s_{\RO}}$ are the three states $|n,n+1\>-|n+1,n\>$, n=0,1,2, one
 obtains:
 $\Psi=|\alpha_0\>\otimes(|1,2\>-|2,1\>)+|\alpha_1>\otimes(|2,0\>-|0,2\>)+
 |\alpha_2\>\otimes(|0,2\>-|2,0\>)$. Since, by virtue of the Schmidt decomposition 
 the three states $|\alpha_n>$ are mutually orthogonal, this is precisely the
 Aharonov state (up to phases that can be changed locally and do not affect the
 correlation in the z-basis). An equivalent test can be performed
 using only local measurements in randomly chosen bases and classical pairewise 
 communication: $\Sen$ choses the bases and $\RO$ announces his results. 

\bibitem{blocking} Similarly to QKD where Eve can block the key distribution, but not
extract information.

\bibitem{FGMR01}
M.~Fitzi, N.~Gisin, U.~Maurer, and O.~von~Rotz.
\newblock Unconditional byzantine agreement and multi-party computation secure
 against dishonest minorities from scratch.
\newblock Manuscript, 2001.

\bibitem{QSSTittel} W. Tittel, N. Gisin and H. Zbinden, Phys. Rev. A {\bf63}, 042301, 2001.

\end{thebibliography}
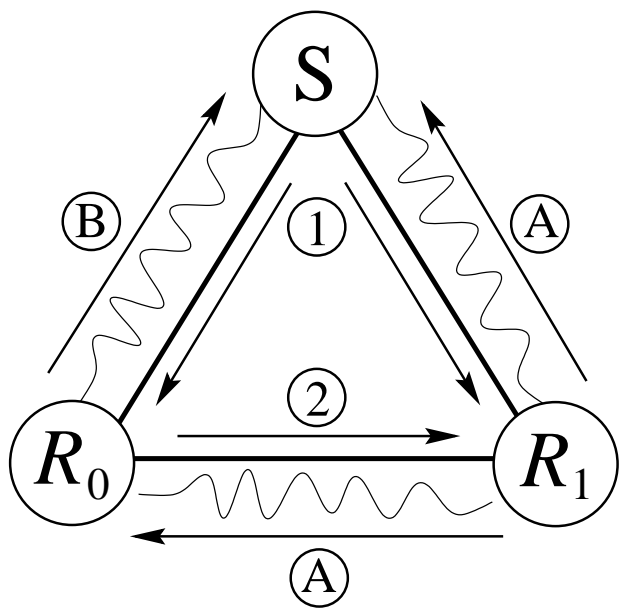
\end{document}